\documentclass[twocolumn,iop,revtex4,usenatbib]{openjournal}
\usepackage[T1]{fontenc}

\usepackage[english, english]{babel}

\DeclareRobustCommand{\VAN}[3]{#2}
\let\VANthebibliography\thebibliography
\def\thebibliography{\DeclareRobustCommand{\VAN}[3]{##3}\VANthebibliography}

\usepackage{algorithm} 
\usepackage[noend]{algpseudocode} 

\usepackage[bitstream-charter]{mathdesign}
\usepackage[usenames]{color}
\usepackage{xcolor}
\usepackage{soul}
\usepackage{multirow}
\usepackage{graphicx}
\usepackage{amsmath}
\usepackage{nicefrac}
\usepackage{microtype}
\usepackage{amssymb}
\usepackage{bbding}
\usepackage{csquotes} 
\usepackage{float}
\usepackage{xurl}
\usepackage{adjustbox}
\usepackage{comment}
\usepackage[hyperfootnotes=false]{hyperref}
\usepackage{tensor} 
\usepackage{enumerate}
\usepackage{orcidlink}
\usepackage[thinlines]{easytable}
\usepackage{booktabs} 
\usepackage{dsfont}

\hypersetup{colorlinks=true, citecolor=blue, linkcolor=blue, urlcolor=blue}
\definecolor{changes}{rgb}{0.4, 0.69, 0.2}
\definecolor{rebeccapurple}{RGB}{102, 51, 153}
\definecolor{apple}{rgb}{0.64, 0.0, 0.0}
\definecolor{darkgreen}{RGB}{77,111,57}
\definecolor{blue(ncs)}{rgb}{0.0, 0.53, 0.74}
\definecolor{cerulean}{rgb}{0.03, 0.27, 0.49}

\DeclareSymbolFont{usualmathcal}{OMS}{cmsy}{m}{n}
\DeclareSymbolFontAlphabet{\mathcal}{usualmathcal}

\definecolor{teal}{RGB}{0,128,128}

\graphicspath{src/macro_sampler/plots}

\newcommand{\ltsima}{$\; \buildrel < \over \sim \;$}
\newcommand{\lsim}{\lower.5ex\hbox{\ltsima}}
\newcommand{\gtsima}{$\; \buildrel > \over \sim \;$}
\newcommand{\gsim}{\lower.5ex\hbox{\gtsima}}
\newcommand{\bra}{\langle}
\newcommand{\ket}{\rangle}

\newcommand{\dd}{\mathrm{d}}

\newcommand{\p}{\mathrm{p}}

\newcommand{\likeli}{\mathcal{L}}

\algdef{SE}[REPEAT]{Repeat}{EndRepeat}[1]{\textbf{repeat} #1 \textbf{times}}{}

\begin{document}

\title{Partition function approach to non-Gaussian likelihoods: \\ information theory and state variables for Bayesian inference\vspace{-30pt}}

\author{Rebecca Maria Kuntz\,\orcidlink{0009-0006-0960-817X}$^{1, \diamond}$}
\author{Heinrich von Campe\,\orcidlink{0009-0006-4071-3576}$^{2}$}
\author{Tobias R{\"o}spel\,\orcidlink{0000-0002-4248-8329}$^{1}$}
\author{Maximilian Philipp Herzog\,\orcidlink{0009-0007-6635-496X}$^{1}$}
\author{Bj{\"o}rn Malte Sch{\"a}fer\,\orcidlink{0000-0002-9453-5772}$^{1,2, \sharp}$}
\thanks{$^\diamond$ \href{mailto:kuntz@stud.uni-heidelberg.de}{kuntz@stud.uni-heidelberg.de}}
\thanks{$^\sharp$ \href{mailto:bjoern.malte.schaefer@uni-heidelberg.de}{bjoern.malte.schaefer@uni-heidelberg.de}}

\affiliation{$^{1}$ Zentrum f{\"u}r Astronomie der Universit{\"a}t Heidelberg, Astronomisches Rechen-Institut,\\ Philosophenweg 12, 69120 Heidelberg, Germany}
\affiliation{$^{2}$ Interdisziplinäres Zentrum f{\"u}r wissenschaftliches Rechnen der Universit{\"a}t Heidelberg,\\ Im Neuenheimer Feld 205, 69120 Heidelberg, Germany; Zuse School ELIZA}

\begin{abstract}
The significance of statistical physics concepts such as entropy extends far beyond classical thermodynamics. We interpret the similarity between partitions in statistical mechanics and partitions in Bayesian inference as an articulation of a result by \citet{jaynes_information_1957}, who clarified that thermodynamics is in essence a theory of information. In this, every sampling process has a mechanical analogue. Consequently, the divide between ensembles of samplers in parameter space and sampling from a mechanical system in thermodynamic equilibrium would be artificial. Based on this realisation, we construct a continuous modelling of a Bayes update akin to a transition between thermodynamic ensembles. This leads to an information theoretic interpretation of Jazinsky's equality, relating the expenditure of work to the influence of data via the likelihood. We propose one way to transfer the vocabulary and the formalism of thermodynamics (energy, work, heat) and statistical mechanics (partition functions) to statistical inference, starting from Bayes' law. Different kinds of inference processes are discussed and relative entropies are shown to follow from suitably constructed partitions as an analytical formulation of sampling processes. Lastly, we propose an effective dimension as a measure of system complexity. A numerical example from cosmology is put forward to illustrate these results. \vspace{+0.2cm}
\end{abstract}

\keywords{inference in cosmology, Monte-Carlo Markov-chains, Bayesian evidence, supernova cosmology}

\maketitle

\section{Introduction}
In problems of statistical inference, Bayes'\@ theorem combines the prior information $\pi(\theta)$ on the  parameters~$\theta \in \mathds{R}^n$ of a physical model with the likelihood $\likeli(y|\theta)$ as the distribution of the data points $y$ for a given parameter choice $\theta$ to the posterior distribution $p(\theta|y)$
\begin{equation} \label{eq:BT}
p(\theta|y) = \frac{\likeli(y|\theta)\pi(\theta)}{p(y)}
\end{equation}
with the Bayesian evidence $p(y)$ given as the normalisation of $p(\theta|y)$ 
\begin{equation}
p(y) = \smallint\dd^n\theta\:\likeli(y|\theta)\pi(\theta). \label{eq: evidence bayes}
\end{equation}
At this instance, one must highlight the fact that use of the integration measure $\smallint \dd^n \theta$ is only justifiable for a flat, Euclidean parameter space with Cartesian coordinates. However, there is no need for us to make any such strong assumption about the geometry of parameter space. In order to consistently define an invariant volume form $\dd \mu(\theta) $, we turn to the field of information geometry.

In this work, we limit ourselves to likelihoods from the so-called exponential family, i.e. the set of probability distributions \citep{amari_information_2016}
\begin{equation}
    \mathcal{P}_e = \{p_\theta(y)\, | \, p_\theta(y) = \exp\left( \theta_i y^i + k(y)-\psi(\theta)\right) \}_{\theta\in \Theta}\,,
 \end{equation}
with $k(y)$ a function of the data $y$, $\psi(\theta)$ ensuring the normalization and $\Theta$ the parameter space (here $\mathds{R}^n$).

Importantly, information geometry tells us that parametric families of probability distributions, such as the exponential family $\mathcal{P}_e$, constitute so-called statistical manifolds, whose metric tensor is precisely the Fisher information $F$, as defined by the likelihood \citep{amari_information_2016, 
nielsen, Nielsen2022TheMF},
\begin{equation}
    F_{\mu\nu}(\theta) = \langle \partial_\mu \ln \mathcal{L}(y | \theta) \partial_\nu \ln \mathcal{L}(y | \theta)\rangle_{y \sim \mathcal{L}}\,. 
\end{equation}
Information geometry further establishes the so-called \mbox{\textit{Jeffreys'\@ prior}} $\pi(\theta)\sim \sqrt{\det F}$, with $F$ the Fisher information, which gives the unique invariant volume form,
\begin{equation}\label{eq: integration measure}
    \dd \mu(\theta)  := \mathrm{d}^n \theta \:\sqrt{\det F}\,,  
 \end{equation}
of a Riemannian statistical manifold, whose metric is $F$ \citep{jeffreysprior, amari_information_2016}. Based on the appearance of the likelihood's Fisher information $F$ in $\pi(\theta)\sim \sqrt{\det F}$, we deem it appropriate to refer to $\sqrt{\det F}$ as \textit{Jeffreys'\@ covolume}, rather than Jeffrey's prior. The volume form in \autoref{eq: integration measure} is invariant under coordinate changes, i.e. under reparameterisations of the statistical model. It correctly reflects the underlying geometry of the parameter space for the family $\mathcal{P}_e$ \citep{jeffreysprior101}.

The structure of Bayes'\@ theorem, with an integral in the denominator and the integrand in the numerator, suggests the definition of a partition function (or \textit{Bayes partition function})  of statistical inference and information theory \citep{partitionfunction101}, 
\begin{equation}
    \label{eq:canZ}
    Z[T,J] = \smallint\dd \mu(\theta)  \:\left[\likeli(y|\theta)\pi(\theta)\:\exp(J_\nu\theta^\nu)\right]^{1/T} \,.
\end{equation}
Clearly, \autoref{eq:canZ} is an extension of the Bayes evidence, modified by the so-called \enquote{sources} $J \in \mathds{R}^n$, as well as a control parameter $T$ in analogy to temperature in statistical physics. This $T$ is in line with the notion of temperature proposed in mathematical optimisation methods, such as simulated annealing \mbox{\citep{Kirkpatrick}}.  For a likelihood $\mathcal{L}(y|\theta) \in \mathcal{P}_e$ together with a prior $\pi(\theta) \sim \exp(\varphi(\theta))$ one finds 
\begin{equation}
    Z[T,J] = \int \dd \mu(\theta)  \:\exp\left(-\frac{1}{T} \left(\Phi(\theta) - J_\alpha\theta^\alpha \right)\right),\label{eq: exponential likeli}
\end{equation} 
allowing to view $\Phi(\theta) := \chi^2(y| \theta)/2 + \varphi(\theta)$ as a potential over parameter space.  Here, $\chi^2(y| \theta)$ includes the statistical model, the data, as well as the data covariance, whilst the function $\varphi(\theta)$ incorporates the prior belief \citep{partitionfunction101}.  Indeed, Markov chain Monte Carlo methods, as introduced to cosmology by \citet{lewis_cosmological_2002}, \, first recognized  the sampler distribution in parameter space as one microscopic realisation of a system subject to~$\Phi$. 

In this picture, an information theoretical partition function as given in \autoref{eq: exponential likeli} governs an ensemble of samplers in the same way that a statistical physics partition would a physical ensemble, completing the correspondence between statistical sampling in inference and their mechanical analogues. Indeed, the Bayes partition marginalises the microscopic realisations of the system, which are subject to randomness, to access its macroscopic features. Only those features at the macroscopic level, e.g. the moments of the posterior distribution, are relevant for inference.  

We consider this great structural similarity to be an articulation of a result by \citet{jaynes_information_1957}, who clarifies that thermodynamics is in essence a theory of information: Inverting Jayne's realisation suggests that statistical mechanics and thermodynamics can be used as tools in information-theoretical problems. Importantly, \autoref{eq: exponential likeli} recovers the Bayesian evidence $p(y)$ for $T = 1$ and $J = 0$. Differentiation of the cumulant generating function $(+ T \ln Z[T,J])$ with respect to $J_\alpha$ gives the cumulants of the posterior distribution $p(\theta|y)$, making them easily accessible at any order \citep{partitionfunction101}. Notably, the source terms $J_\alpha$, which appear alongside an inverse temperature scaling, receive a mechanical intuition when a physics perspective is adopted by way of the potential $\Phi(\theta)$. In this, $J_\alpha$ appears as an additional control variable alongside $T$ which naively introduces a shift to the potential $\Phi(\theta)$ in parameter space, thus causing a non-trivial alteration of the sampler configuration in parameter space. Accordingly, we refer to $\Phi_J(\theta) := \Phi(\theta)-J_\alpha\theta^\alpha$ as the shifted potential. \newline 

To begin with, Sect.~\ref{sect_parametric_transition} presents a continuous parametrization of a Bayes update, in order to continuously quantify the intermediate stages of a belief update. In this, a variant of the Jarzynski equality for transitions between equilibrium ensembles under expenditure of mechanical work is transferred to statistical inference. Subsequently, Sect.~\ref{sect_differential} furthers the transfer of concepts from thermodynamics to information theory by arranging the structure of Bayesian inference in parallel to that of thermodynamics. 
After this, Sect.~\ref{sect_reversible_processes} considers statistical inference processes against the backdrop of the first law of thermodynamics.  Sect.~\ref{sect: macrocanonical_ensemble} generalizes this approach to a setting where the sampler number $N$ can vary. Besides, Sect.~\ref{sect_rel_entropies_from_bayes_partitions} seeks to generalise the link between relative entropies and partitions in information theory.
Finally, Sect.~\ref{sect_effective_dimension} presents the concept of an effective dimension as a temperature-dependent measure of system complexity. Sect.~\ref{sect_cosmology_application} illustrates the thermodynamic viewpoint using an inference problem from cosmology (placing joint constraints on the matter density $\Omega_m$, and the dark energy equation of state parameter~$w_0$). 
Sect.~\ref{sect_summary} summarizes our results.

Throughout the paper, we use the summation convention. For parameter tuples $\theta \in \mathds{R}^n$ and data tuples $y \in \mathds{R}^m$ as vectors with contravariant indices; Greek indices are reserved for quantities in parameter space ($\theta^\alpha$), while Latin indices denote objects in data space ($y^i$).

\section{Parametric transition from prior to posterior}
\label{sect_parametric_transition}

We begin by viewing a Bayes update in analogy to a continuous transition between two equilibrium ensembles in thermodynamics. To achieve this, a control parameter $\lambda$ is introduced to the Bayes-partition (setting $J=0$), 
\begin{equation}
    Z_\lambda[T] = \smallint \dd \mu(\theta)  \: [\likeli(y|\theta)^\lambda \pi(\theta)]^\frac{1}{T}.
    \label{eq: introduce lambda}
\end{equation}
For conciseness of notation, we adopt the shorthand $\likeli(y|\theta)^\lambda \pi(\theta) \to \likeli^\lambda \pi$ in this section. Effectively, $\lambda$ is an additional state variable which governs the influence of the data on the inference process.

At the same time, \autoref{eq: introduce lambda} allows to quantify the rate of change in Shannon's information entropy $S$ with progressing inference \mbox{($\lambda: 0 \to 1$)}. Effectively, $\lambda$ describes how the conversion from prior $\pi$ to posterior belief occurs through the influence of the likelihood $\mathcal{L}$ at fixed $T, J$. In thermodynamics, the information entropy $S$ directly follows from the partition function $Z$ via $S(T, (\dots)) = \partial_T (T \ln Z [T,(\dots)])$. Here, $(\dots)$ indicates other control variables, which are not specified further at this point.

In the information theoretic case at hand, the \emph{information} entropy $S$ is derived from $Z$ analogously, leading to   
\begin{align}
    S_\lambda(T) &=\ln  Z_\lambda[T] -  (T \: Z_\lambda[T])^{-1} \smallint \dd \mu(\theta)  \; \left(\likeli^\lambda \pi\right)^{1/T} \ln \left(\likeli^\lambda \pi\right)  \,,
\end{align}
where the total entropy difference $\Delta S_{\pi \to \lambda}$ between prior and the \enquote{intermediate distribution} $\likeli^\lambda \pi$ at stage $\lambda\neq0$ of the inference progression can be written as  
\begin{align}
    \Delta S_{\pi \to \lambda} (T) &:=  S_{\pi}(T) - S_{\lambda}(T) \\  &= \ln \smallint \dd \mu(\theta)  \: (\pi^{1/T} / Z_\lambda[T])+ \nonumber \\ &\quad + (T \: Z_\lambda[T])^{-1} \smallint \dd \mu(\theta)  \; \left(\likeli^\lambda \pi\right)^{1/T} \ln \left(\likeli^\lambda \pi\right) \nonumber \\ &\quad- \left(T \: \smallint \dd \mu(\theta)  \: \pi^{1/T}\right)^{-1}\smallint \dd \mu(\theta)  \: \pi^{1/T} \ln \pi\,. 
\end{align}
Intuitively, this way of introducing $\lambda$ signifies a weighting of the inverse data covariance $C$. This is made plain for a $\lambda$-scaled exponential likelihood,
\begin{equation}
    \mathcal{L}^\lambda \sim  \exp\left(-\frac{1}{2}\left(y^i - y^i_{\text{model}}\right)\;\lambda C_{ij} \;\left(y^j - y^j_{\text{model}}\right)\right)\,. 
\end{equation}
In the most illustrative case of uncorrelated data, \mbox{$C_{ij} = 0 \;\forall i\neq j$} and $C_{ii} = (\sigma_i \sigma_i)^{-1}$, $\lambda$ acts as a weighting for the data variance. Here, the case of $\lambda = 0$ (i.e. only the prior is known) corresponds to 
\begin{equation}
    (\sigma_e^{i} \sigma_e^{j})^{-1}:= \lambda (\sigma^{i}\sigma^j)^{-1} = 0 \; \leftrightarrow \; \sigma_e^{i} \sigma_e^{j} = \infty\label{eq: lambda scaling}
\end{equation} 
before measurement. Not having measured the data yet is equivalent to having infinite \textit{effective} uncertainty $(\sigma_e^{i} \sigma_e^{j})^{-1}$ in the data. Likewise, for $\lambda = 1$, the effective uncertainties coincide with the \textit{actual} uncertainties of the data set, i,e, $\sigma_e^{i} \sigma_e^{j} = \sigma^{i}\sigma^j$. In conclusion, $\lambda$ interpolates between constraining data with lower experimental errors ($\lambda=1$) and uninformative data with effectivelty infinite experimental errors ($\lambda=0$). Choosing a $\lambda$-scaled variant of \autoref{eq: exponential likeli} at $J=0$, i.e. 
\begin{equation}
    Z_\lambda[T] = \int \dd \mu(\theta)  \; \exp\left(-\frac{1}{T}\left(\lambda (\chi^2/2)+\varphi(\theta)\right)\right) \,
\end{equation}
with a prior $\pi(\theta)\sim\exp(-\varphi(\theta))$, one finds 
\begin{align}
    \mathrm{d}S\big|_{\substack{T=1 \\ J =0}} &= \left(\frac{\partial S}{\partial \lambda}\right)\: \mathrm{d}\lambda = \left(-\lambda \text{Var}(\chi^2) - \text{Cov}(\chi^2, \varphi)\right)\:\mathrm{d}\lambda\,, 
    \label{information entropy decrease with lambda}
\end{align}
with $\dd S$ the rate of change of uncertainty while incorporating new information. This allows for two important observations: Firstly, the uncertainty strictly decreases as inference progresses, and secondly, the rate of information gain with inference depends on the variance of the likelihood's potential $\text{Var}(\chi^2)$ and the covariance between the likelihood and prior potentials $\text{Cov}(\chi^2, \varphi_0)$.

 At this point, the relative entropy, (the Kullback-Leibler (KL) divergence) between prior and posterior is discussed as a function of $\lambda$. The KL-divergence between two probability distributions $p,q$ over parameter space, 
\begin{equation}
    \text{D}_{KL}[p(\theta)\, ,\, q(\theta)] = \int \dd \mu(\theta)  \; p(\theta) \ln \left(\frac{p(\theta)}{q(\theta)}\right)\,,
\end{equation}
is an oriented measure of their relative information entropy \citep{Nielsen2022TheMF}. 
One finds 
\begin{align}
    \mathrm{D}_{KL}[p(\theta, \lambda| y),\;\pi(\theta)] 
    = p(y)^{-1} \smallint \dd \mu(\theta)  \: \mathcal{L}^\lambda \pi \; \ln \mathcal{L^\lambda} - \ln p(y) \,.
\end{align}
This is in accordance with expectations, since the limiting cases are
\begin{equation}
    \mathrm{D}_{KL}[p(\theta, \lambda| y),\;\pi(\theta)] \; \stackrel{\lambda \to 0}{\longrightarrow}\; 0 
\end{equation}
i.e. the prior coincides with the intermediate distribution when no measurement has been taken yet. For $\lambda\to 1$, i.e. completion of a full update step, the KL-divergence between posterior and prior
\begin{equation}
    \mathrm{D}_{KL}[p(\theta, \lambda| y),\;\pi(\theta)] \; \stackrel{\lambda \to 1}{\longrightarrow}\;\int \dd \mu(\theta)  \: p(\theta|y) \; \ln \left(\frac{p(\theta|y)}{\pi(\theta)}\right) \,
\end{equation}
is recovered. This is also referred to as the \emph{surprise statistic} \citep{pedromiguelbenni}. Here, $(\mathrm{d}\mathrm{D}_{KL}/\mathrm{d}\lambda)$ is the rate of information increase with progressing inference, which could be useful in experimental design.

\subsection{Jarzynski equality}
Following \citet{Jarzynski_1997}, the average work $W$ expended in a infinitely fast switching between two equilibrium states of a thermodynamic ensemble is given by
\begin{equation}
    \langle e^{- W/T}\rangle_0 = e^{-\Delta G/T} \,,
    \label{eq: work gibbs relation}
\end{equation} 
with $\langle \cdot \rangle_0$ the equilibrium average at $\lambda = 0$, and $G$ the Gibbs free energy \citep[see also][]{Crooks1998NonequilibriumMO}. Importantly, \autoref{eq: work gibbs relation} assumes that the transition occurs through a series of equilibrium states. For \mbox{$\lambda:\: 0 \to 1$}, $W$ corresponds to the average  \enquote{work}  required for a full inference step. The information theoretic analogue of the potential $G$ is given by $G(T,\lambda) = - T \ln Z_\lambda[T,\lambda]$. For a detailed discussion of this identification, the reader is referred to Sect.~\ref{sect_differential}. With this choice, we find 
\begin{align}
    \Delta G(T, \lambda) &= G(T, \lambda) - G(T, 0) \\
    &= -T \ln Z_\lambda[T] + T \ln Z_0[T] \\
    &= T \ln \frac{\int_K  \dd \mu(\theta)  \: \pi^{1/T}}{\int_K \dd \mu(\theta)  \: \left( \mathcal{L}^\lambda \pi\right)^{1/T}}\, ,
\end{align}
with $J=0$. For a parameter domain $K\subseteq \mathds{R}^n $, \autoref{eq: work gibbs relation} leads to 
\begin{align}
    \frac{\int_K \dd \mu(\theta)  \left[\pi \: e^{- W} \right]^{1/T}}{\int_K \dd \mu(\theta)  \: \pi^{1/T}} &= \frac{\int_K \dd \mu(\theta)  \: \left[ \mathcal{L}^\lambda \pi\right]^{1/T}}{\int_K \dd \mu(\theta)  \: \pi^{1/T}} \, , \label{eq: equality for all K}
\end{align}
which ultimately results in
\begin{equation}
    e^{- W} = \mathcal{L}^\lambda \; \leftrightarrow \; W = - \lambda \ln \mathcal{L} = \lambda\: \frac{\chi^2}{2} \,, \label{eq: work and likelih}
\end{equation}
since the equality of the integrals in \autoref{eq: equality for all K} holds for any domain of integration $K$.
\autoref{eq: work and likelih} is highly intuitive and consistent with \citet{Dahlsten_2011}: The amount of \enquote{work} $W$  expended in inference equals the effective potential energy contributed by the likelihood at the intermediate stage $\lambda$ of the update. In that sense, the work performed to update a parameter space ensemble in light of new data depends on the data itself as well as the model choice via the likelihood.

The illustrative role played by the analogue of mechanical work $W$ in statistical inference (\autoref{eq: work and likelih}) motivates Sect.~\ref{sect_differential}, which aims to transfer thermodynamical concepts to information theory.

\section{Thermodynamic perspective on Bayes partitions}\label{sect_differential}
The Guggenheim scheme (on the left side of \autoref{fig: Guggenheim scheme information theory}) is commonly used to organize (extensive and intensive) thermodynamic quantities, so as to visualise their relations, e.g. Maxwell's relations. At fixed particle number $N$, \autoref{fig: Guggenheim scheme information theory} shows the extensive internal energy $U(S,V,N)$, which is a function of the thermodynamic entropy $S$, as well as the physical volume $V$, which are both extensive quantities. The respective opposing corners of the Guggenheim scheme give the conjugate intensive quantities of $S$ and $V$, namely the temperature $T$ and the pressure $p$, respectively. Those intensive state variables define the Gibbs free energy $G$ (together with the extensive particle number $N$) via a Legendre transformation. Moreover, the Guggenheim scheme on the left side of \autoref{fig: Guggenheim scheme information theory} includes the Helmholtz free energy $F(T,V,N)$ as well as the Gibbs enthalpy $H(S,p,N)$.

At this point, the structural counterparts of information theoretical quantities in thermodynamics must be identified. One way to do this is to locate the controlled parameters of sampler ensembles in a scheme akin to \autoref{fig: Guggenheim scheme information theory}. This is done using the concept of homogeneity in thermodynamics. 

As of now, we limit ourselves to a Gaussian likelihood  
\begin{equation}
    \mathcal{L}(y| \theta) \sim \exp\left(-\frac{1}{2T} F_{\mu\nu}\theta^\mu \theta^\nu\right)\,,  \label{eq: Gaussian likeli}
\end{equation}
governing a sampler ensemble in parameter space with a parabolic log-likelihood $\propto F_{\mu\nu}\theta^\mu \theta^\nu$. The Bayes partition (see \autoref{eq:canZ}) further entails source terms $J_\alpha$, as well as a prior, which we chose to be uninformative, e.g. $\pi(\theta) \sim \mathrm{const}$. 

Note that the covolume $\sqrt{\det F}$ doubles as a diagnostic tool for the
degeneracy of an inference problem as the following example demonstrates:
\cite{partitionfunction101} derive the Fisher information of a linear model
$y^i = A\indices{^i_\mu} \theta^\mu$ (i.e. a Gaussian likelihood) as 
\begin{equation}
    F_{\mu\nu} = C_{ij} A\indices{^i_\mu}  A\indices{^j_\nu} ,
\end{equation}
where $C_{ij}$ is the covariance matrix of the $N$ datapoints.  Computation of its determinant is straightforward, 
\begin{equation}
\det F = \varepsilon^{\mu_1 \ldots \mu_n} C_{i_1 j_1} A\indices{^{i_1}_{1}} A\indices{^{j_1}_{\mu_1}}
\ldots C_{i_n j_n} A\indices{^{i_n}_{n}} A\indices{^{j_n}_{\mu_n}}\,,
\end{equation}
with $\varepsilon^{\mu_1 \ldots \mu_n}$ the totally antisymmetric Levi-Civita symbol in $n$ dimensions. 
For a non-vanishing determinant, the values of the $j$'s must be
pairwise distinct

in some terms of the sum. In all others, there would be at least one pair 
of $A\indices{^j_\mu}$'s with equal $j$'s, which would yield zero when contracted with 
the antisymmetric Levi-Civita symbol.

However, for an underconstrained model, i.e. if there are fewer data points than model parameters, $N < n$, the $j$'s cannot be
chosen in a pairwise-distinct way in any of the sum's terms. Thus, $\det F$ must vanish for an
underconstrained model. This result is intuitive since a vanishing covolume
signifies that the $N$ datapoints at hand are not sufficient two distinguish
between different choices of the $n$ parameters.

The partition function  for a Gaussian likelihood governing a single sampler in parameter space can be written as
\begin{align}
    Z [T,J] &= \int \dd \mu(\theta)  \exp\left(-\frac{1}{2T} F_{\mu \nu}\theta^\mu \theta^\nu + \frac{1}{T} J_\alpha \theta^\alpha\right)\nonumber \\
    &= \sqrt{(2\pi T)^n}\;\exp \left(+\frac{1}{2T} F^{\alpha \beta} J_\alpha J_\beta\right)\,,\label{eq: Gaussian partition}
\end{align}
using the analytic solution of a Gaussian integral.
The construction of a conventional thermodynamic system requires its extensive variables to be related via a consistent fundamental relation. Importantly, the Gibbs-Duhem relation forbids any thermodynamic potential that only depends on intensive state variables, as information about quantities like total energy would otherwise be inaccessible. Intuitively, extensive variables scale with system size. However, for applications in pure statistics, concepts such as system size, extensivity and intensivity of state variables require a case-by-case definition. Therefore, after selecting extensive variables, we must check that their interplay is consistent.

\begin{figure}
    \centering
    \includegraphics[width=.45\textwidth]{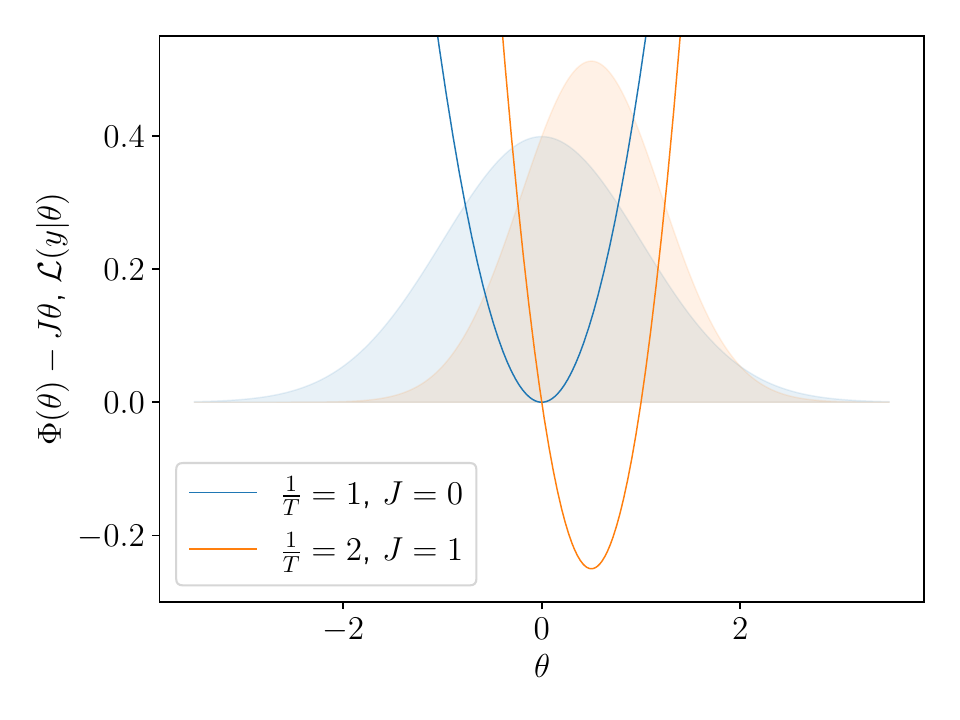}
    \caption{Schematic changes of a univariate Gaussian (shaded region) and its corresponding potential (solid line), as given in equation \eqref{eq: Gaussian partition}, with different values of $T$ and $J$. For simplicity, $F = 1$ is fixed.}
    \label{fig:beta_j_dependence}
\end{figure}

In an inference setting, the quantities $T$ and $J_\alpha$ function as external control variables. Their influence on the potential and the likelihood is depicted schematically
in \autoref{fig:beta_j_dependence}. While $T$ effectively scales the
covariance and thus the width of the Gaussian, $J_\alpha$ induces a shift in the governing potential.
We do not expect $J_\alpha$ to scale with what we choose to see as the system size (the extent of the sampler ensemble in parameter space). Thus, $J_\alpha$ is chosen to be a candidate for an intensive quantity. Naturally, the information theoretic variants of $U,S,N$ are viewed as extensive variables. Importantly, first-order homogeneity of the fundamental relation necessitates another identification: $n = \text{dim}\:\Theta$ (with $\Theta$ the parameter space) is selected as an extensive variable, changes of which will be ignored (set $\dd n = 0$ always) in all standard inference settings. 

From a statistical point of view, changes in $n$ would signify a different number of parameters and are thus related to the problem of model selection as discussed in \citet{schosser2024markovwalkexplorationmodel}. For a thermodynamic viewpoint, $n$ will turn out to mimic the notion of degrees of freedom (DOF) in physics. Furthermore, it will be shown that the information entropy, which is a measure of system complexity, scales with $n$, suggesting its extensive character. In fact, the link between $n$ and system complexity will be the basis of the effective dimension presented in Sect.~\ref{sect_effective_dimension}. Now, we must check first-order homogeneity of the fundamental relation in all extensive quantities.

With these choices in place, the potential $G(T,J_\alpha,n) :=  - T \ln Z[T,J_\alpha,n]$ (for one sampler) parallels the structure of the Gibbs free energy~\mbox{$G(T,p,N)= - T \ln Z[T,p,N]$},  
\begin{align}
    G(T,p,N) &= U - TS + pV \ (+\mu N)\,,\label{eq: gibbs free energy and Z}\\
    \dd G(T,p,N) &= - S\dd T + V\dd p\; (+\mu \dd N)\,.
\end{align}
Now, the conjugate quantities of all thermodynamic-like variables must be derived. The conjugate quantity of $J$ follows as
\begin{equation}
    \partial_{J_\alpha} \left(-T \ln Z[T,J,N=1,n] \right)= - F^{\alpha \beta}J_\beta\,. 
\end{equation}
Accordingly, the conjugate quantity of $n$, namely $\eta := (T/2)\ln (2\pi T)$, is clearly intensive. Now, we make use of the following shorthand notation
\begin{align}
    J_\alpha \hat{=} \:J,\;\;
    J_\alpha J_\beta \hat{=}\: JJ,\;\;  F_{\mu \nu} \hat{=}\: F,\;\; F^{\mu \nu} \hat{=}\: F^{-1},
\end{align}
for simplicity. The formulation of all results in index notation is straightforward. Setting $\dd n = 0,\:\dd N = 0$ (and $\mu = 0$) in the following, one ultimately finds 
\begin{align}
    G(T,J,n) &= U-TS - \frac{1}{2}F^{-1}JJ - n \eta \\ \dd G &= - S \dd T - \frac{1}{2} F^{-1}\dd (JJ)\,.\label{eq: gibbs again}
\end{align} 
By looking at \eqref{eq: gibbs free energy and Z}, one can directly perform the intensity extensivity check:
since the two thermodynamic potentials $G$ and $U$ are both extensive, the other terms must contain one intensive and one extensive variable each (as in the definition of the \emph{thermodynamic} Gibbs free energy). This confirms that $JJ$ can structurally assume the role of an intensive state variable. The converse argument shows that $F^{-1}$ can indeed be extensive. On the basis of these findings, we make an adjustment to the selected variables. Given that the parameter covariance $F^{-1}$ characterizes the extent of a region in $\Theta$ where samples lie (unlike a volume, this does not impose rigid boundaries), its extensivity seems to be favorable. Consequently, we discuss $F^{-1}\dd (JJ) (= 2 F^{-1}J\dd J)$ rather than $F^{-1}J\dd J$.  Now, it is structurally well-justified (omitting the $n$- and $N$-direction of the diagram) to build an information theoretic Guggenheim scheme for a Gaussian likelihood, as given in
\autoref{fig: Guggenheim scheme information theory}.

\begin{figure}
    \begin{minipage}[t]{0.48\columnwidth}
        \centering 
        \begin{TAB}(e,1.4cm,1.4cm){|c|c|c|}{|c|c|c|}
             \color{apple} $-S$ & $\color{apple}\mathbf{U}(S, V)$ &  \color{apple} $V$\\
            $\color{apple}\mathbf{H}(S,\color{cerulean} p)$ &  & $\color{apple}\mathbf{F}(\color{cerulean}T, \color{apple}V)$ \\
           \color{cerulean} $- p $ & $\color{apple}\mathbf{G}(\color{cerulean}T, p)$ & $\color{cerulean}T$ 
        \end{TAB}
    \end{minipage}
    \hfill
    \begin{minipage}[t]{0.48\columnwidth}
        \centering
        \begin{TAB}(e,1.4cm,1.4cm){|c|c|c|}{|c|c|c|}
            $\color{apple}-S$ & $\color{apple}\mathbf{U}(S, F^{-1})$ &  $\color{apple} F^{-1}$\\
            $\color{apple}\mathbf{H}(S, \color{cerulean}JJ)$ &  & $\color{apple}\mathbf{F}(\color{cerulean}T,\color{apple} F^{-1})$\\
            $\color{cerulean}\frac{1}{2}JJ $ &  $\color{apple}\mathbf{G}(\color{cerulean}T, JJ)$ & $\color{cerulean}T$
        \end{TAB}
    \end{minipage}
    \caption{The Guggenheim scheme of standard thermodynamics (at fixed $N$), on the left, and one possibility to construct the analogous Guggenheim scheme of information theory for a Gaussian likelihood (at fixed $n$ and $N$), on the right. \color{cerulean}Intensive variables in blue, \color{apple} extensive variables in red. \newline }
    \label{fig: Guggenheim scheme information theory}
\end{figure}

\subsection{Derived quantities from Bayes partitions}
Within this framework, \mbox{\autoref{eq:canZ}} is employed to derive counterparts of thermodynamic quantities. In the following, the expectation value of a random variable $X$ is understood to be 
\begin{equation}
    \bra X \ket =  Z [T, J_\alpha]^{-1} \int\dd\mu\; X \exp\left(-\frac{1}{T}\left( \Phi (\theta)-J_\alpha\theta^\alpha\right)\right)\,. 
\end{equation} 
In a Gibbs ensemble, the internal energy $U$, the enthalpy $H$ and the entropy $S$ of the parameter space configuration follow from the partition $Z[T,p] := Z[T,p,N=1]$ as
\begin{align}
    H(S,p) &= U + pV = T^2 \partial_T \mathrm{ln} Z [T,p] \label{eq: enthalpy} \\
    U(S, V) &= T^2 \partial_T \mathrm{ln} Z [T,p] \label{eq: U from F} - pV \, ,\\
    S(T, p) &= \partial_T (T \ln Z [T,p]),\, \label{eq: entropy}\,. 
\end{align}
Analogously, for the previous identifications, we find (with $Z[T,J] := Z[T,J, n=\text{const},N=1]$), 
\begin{align}
    H(S,J) & = T^2 \partial_T \mathrm{ln} Z [T,J] \label{eq: enthalpy} \\
    U(S, F^{-1}) &= T^2 \partial_T \mathrm{ln} Z [T,J]\label{eq: U from F} + \frac{1}{2} F^{-1}JJ\, ,\\
    S(T, J) &= \partial_T (T \ln Z [T,J])\, \label{eq:entropy_j}\,. 
\end{align}
The result for the enthalpy $H$ in \autoref{eq: enthalpy} coincides with the expectation value of the shifted potential \mbox{$\langle \Phi_J \rangle = T^2 \partial_T \mathrm{ln} Z[T,J,N]$}, which follows directly using
\autoref{eq: exponential likeli}.
Employing \autoref{eq: Gaussian partition}, we find
\begin{align}
    H &= T^2\partial_T \ln Z[T, J] = \frac{nT}{2}-\frac{1}{2} F^{-1} JJ\:, \label{eq: H for Gauss}\\
    U &= T^2\partial_T \ln Z[T, J] + \frac{1}{2} F^{-1} JJ = \frac{nT}{2}\label{eq: U for Gauss}\:,\\
    S &= \ln Z[T, J] + T \partial_T \ln Z = \frac{n}{2}\left(\ln 2\pi T + 1\right)\:.
\end{align}
Clearly, the fact that \mbox{$S \sim n \times h(T)$} (with $h$ some function only dependent on the intensive $T$) confirms the extensivity of $n$ in the information theoretical complex displayed in \autoref{fig: Guggenheim scheme information theory}. Interestingly, the result for $U$ in \autoref{eq: U for Gauss} 
is reminiscent of the internal energy \mbox{$ U = f k_B T/2$} of a one-particle system in thermal equilibrium (with $f$ the number of translational DOFs and $k_B$ the Boltzmann constant). For $J\to 0:$ \mbox{$H \to \langle \Phi \rangle = (nT/2)\; (= U)$}.

The total differentials of $S$, $H$ and $U$ for \autoref{eq: Gaussian partition} turn out to be 
\begin{align}
    \dd H = &\left(\frac{\partial H}{\partial S}\right)_{JJ} \dd S + \left(\frac{\partial H}{\partial (JJ)}\right)_S \dd (JJ) \\
    = &\frac{n}{2} \; \dd T - \frac{1}{2}F^{-1} \;\dd (JJ) \label{equ:dH} \\
    \label{equ:dU}
    \dd U = &\left(\frac{\partial U}{\partial T}\right)_{F^{-1}} \dd T + \left(\frac{\partial U}{\partial F^{-1}}\right)_T \dd F^{-1} = \frac{n}{2}.\\
    \dd S = &\left(\frac{\partial S}{\partial T}\right)_{JJ} \dd T + \left(\frac{\partial S}{\partial (JJ)}\right)_T \dd (JJ)
    = \frac{n}{2T}\:\dd T \,. \label{equ:dS} 
\end{align}
Changes in $H$ and $S$ with respect to temperature changes scale with the variance of the (shifted) potential $\text{Var}(\Phi_J)_{\text{Gauss}} \;(= \text{Var}(\Phi)_{\text{Gauss}})$, which is defined by the parameter space dimension $n$ (as well as $T$) in the Gaussian case, i.e. 
\begin{equation}
    \text{Var}(\Phi_J)_{\text{Gauss}} = \frac{nT^2}{2} \;\left( = \text{Var}(\Phi)_{\text{Gauss}}\right).
\end{equation}
This is in line with expectations, as temperature changes modify the size of confidence regions in parameter space by altering the width of the governing potential. This concept is used to increase sampling efficiency in methods such as simulated annealing by accelerating the convergence of samplers into the parameter space potential's minima \citep{Kirkpatrick}. Moreover, the exclusive appearance of $\langle\theta\rangle$ in the derivatives with respect to $J$ alludes to the ability of the "shift" in $\Phi_J$ to relocate the maximum of the posterior in parameter space.

Use of \autoref{equ:dU} yields the caloric equation of state analogue, 
\begin{align}
    \dd U = \left( \frac{\partial U}{\partial T} \right)_{F^{\alpha \beta}, J} \:\dd T 
    = \frac{n}{2}\: \dd T\:.
\end{align}
At this instance, the statistics equivalent of a thermodynamic heat capacity is inspected. In physics, the heat capacity at constant pressure or volume ($C_p$ and $C_V$, respectively) measures the amount of thermal energy $\delta Q$ necessary to foster a certain temperature change $\dd T$ of the system. In thermodynamics, one defines 
\begin{equation}
     C_p := \left(\frac{\partial H}{\partial T} \right)_p \quad \text{and}\quad C_V := \left( \frac{\partial U}{\partial T} \right)_V \,.
\end{equation}
For an ensemble of samplers subject to a likelihood as given in \autoref{eq: Gaussian likeli}, the resulting heat capacity, 
\begin{equation}
    C_{JJ} =  C_{F^{-1}} = \frac{n}{2}\,, \label{eq: heat gaussian}
\end{equation}
is once again reminiscent of the scaling of $C_p$ with the number of internal DOFs $f$ of a monoatomic gas. Most importantly, $C_p$ and $C_V$ are always extensive. Consequently, \autoref{eq: heat gaussian} confirms the previous assumption of $n$ being extensive in this structure.

Lastly, please note that the generalization of these results to a non-Gaussian
setting is highly non-trivial. In particular, the identification of conjugate variables can only ever be possible 
when there is a close form expression for derivatives of \autoref{eq:canZ}.

\section{Reversible Processes in Statistical Inference}
\label{sect_reversible_processes}

The thermodynamic perspective on statistical inference motivates the interpretation of Bayes-updates in analogy to thermodynamic processes. The first law of thermodynamics $\dd U = \delta W + \delta Q $ describes the change in a system's internal energy $U$ as the sum of contributions from work/mechanical energy $\delta W$ as well as heat transfer/thermal energy $\delta Q$. Of course, in an information theoretic picture, any physics analogy has limitations: We note that in the statistics setting at hand, the enthalpy-analogue $H$ (see \autoref{eq: H for Gauss}) structurally assumes the role that is played by the internal energy $U$ in a canonical thermodynamic ensemble. Based on this, this section discusses the splitting of $\dd H$ in information theoretic variants of work $\delta W$ and heat $\delta Q$, respectively. The anatomy presented in \autoref{fig: Guggenheim scheme information theory} is employed, keeping in mind that the Bayes partition in \autoref{eq: Gaussian partition} is representative of a Gibbs-like ensemble governed by a quadratic potential. Once again, the alteration of the sampler configuration in light of new data is understood in parallel to a transition between two states of a thermodynamic system. These observations  could potentially offer a novel  view on different sampling modes in the context of Markov chain Monte Carlo methods.  However, more extensive investigation is needed to make this connection plain.

\subsection*{\enquote{Adiabatic} process}
To begin with, it is instructive to discuss the analogue of a reversible adiabatic process. A process
that is both reversible and adiabatic is called isentropic. In thermodynamics, a process is called
\textit{reversible} if it can be undone without lasting change to the environment.
Crucially, reversibility does not imply adiabaticity and vice versa.  For an isentropic process the change in $H$ is constituted solely by the work expended in the transition,
i.e.
\begin{equation}
    \dd H = \dd W \quad \text{and} \quad \delta Q = T \dd S \stackrel{!}{=}0 \,.
    \label{eq: isentropic dU}
\end{equation}
Consider an isentropic process, brought about by a change $\dd F^{-1} \neq 0$ with $\dd (JJ) = 0$. Evidently, the true (likelihood) posterior is only recovered when we set $J=0$ (before) after statistical inference. Accordingly, an isochoric transition at \textit{zero} potential shift would enforce faithfulness to the true distributions at every step of the inference process (which is a stronger constraint than the usual $J_{\text{init.}} = 0 \; \wedge\; J_{\text{fin.}} = 0$). For an isentropic process, one finds
\begin{equation}
    \dd T = - \left(\frac{\partial S}{\partial T}\right)^{-1} \left(\frac{\partial S}{\partial F^{-1}}\right) \: \dd F^{-1} \,. 
    \label{eq: isentropic dS}
\end{equation}
\noindent 
Combining \autoref{eq: isentropic dU} and \autoref{eq: isentropic dS} results in 
\begin{equation}
    \dd H = \dd W = 
    \left(\left(\frac{\partial H}{\partial F^{-1}}\right) - \left(\frac{\partial H}{\partial T}\right) \left(\frac{\partial S}{\partial T}\right)^{-1} \left(\frac{\partial S}{\partial F^{-1}}\right) \right)\:\dd F^{-1}. 
\end{equation}
For \autoref{eq: H for Gauss}, this turns out to be 
\begin{align}
    \dd H = \dd W = - \frac{1}{2} F^{-1} \dd (JJ) (= - F^{-1} J \dd J), 
\end{align}
whose structure is reminiscent of the thermodynamic \textit{canonical} result $\dd U = \dd W = + p\dd V$. Notably, the sign difference is ultimately due to the different way in which $J$ appears in the Bayes partition: $\exp(-J \theta)$ rather than $\exp(+pV)$ in thermodynamics.

\subsection*{\enquote{Isothermal} process}
Secondly, a reversible isothermal process ($\dd T = 0, \: \dd F^{-1} \neq 0, \:\dd (JJ) \neq 0$) is discussed. Similarly, in  thermodynamics, such a process articulates as a change in both $p$ and $V$. From a Bayesian point of view, any inference process only results in the true posterior when $T=1$ is restored or fixed throughout sampling. For a reversible process, the 2nd law of thermodynamics implies that $\delta Q = T \dd S$. For the reversible isothermal case one finds
\begin{align}
    \dd H 
    &\stackrel{\text{rev.}}{=}  T \dd S + \delta W \\ &\stackrel{(\dd T = 0)}{=} T\left( \left(\frac{\partial S}{\partial F^{-1}}\right)_T \:\dd F^{-1}+  \left(\frac{\partial S}{\partial J}\right)_T \:\dd J\right) + \delta W \, , 
\end{align}
which results in 
\begin{align}
    \dd H = \dd W = -\left(\frac{1}{2}F^{-1} \;\dd (JJ) + \frac{1}{2} JJ \dd F^{-1}\right)\,
\end{align}
for a reversible isothermal process in an open system. This result resembles the internal energy change for an open canonical system $\dd U = p \dd V + V \dd p$ in thermodynamics. Clearly, $p$ and $V$ can both change independently only if one allows for an open system.

\subsection*{\enquote{Isobaric} process}
\noindent Consider a process occurring under $\dd(JJ) = 0$. Then 
\begin{align}
    \dd H &\stackrel{(\dd F^{-1} = 0)}{=} \left(\frac{\partial H}{\partial T}\right)_{JJ,F^{-1}}\:\dd T + \left(\frac{\partial H}{\partial JJ}\right)_{T,F^{-1}} \: \dd (JJ)\\ &= \delta Q + \delta W.  
\end{align}
This leads to 
\begin{align}
    \delta Q &=  \frac{n}{2} \:\dd T = C_{JJ} \:\dd T\quad \text{and}\quad \label{eq: heat isobaric}\\
    \delta W &= \dd H -  \frac{n}{2} \:\dd T = 0.\label{eq: work isobaric}
\end{align}
In a thermodynamic Gibbs ensemble, the pressure $p$ structurally takes the role that the volume $V$ has in a canonical ensemble. Thus, the structure of an "isobaric" process in this information theoretic Gibbs-like ensemble is comparable to the structure of the thermodynamic \emph{isochoric} process in a canonical ensemble.  
Such an isochoric process performed on an ideal gas (a canonical ensemble) is characterized by a change $\dd U = \dd Q = C_V \:\dd T$ and $\dd W = 0$, which evidently parallels \autoref{eq: heat isobaric} and \autoref{eq: work isobaric}.

\subsection*{\enquote{Isochoric} process}
At this point, the analysis turns to a  reversible isochoric process ($\dd F^{-1} = 0$), where the change in the system's state is brought about by $\dd T \neq 0$. In this case, 
\begin{align}
    \dd H &\stackrel{(\dd F^{-1} = 0)}{=} \left(\frac{\partial H}{\partial T}\right)_{JJ,F^{-1}}\:\dd T + \left(\frac{\partial H}{\partial JJ}\right)_{T,F^{-1}} \: \dd (JJ)\\ &= \delta Q + \delta W. 
\end{align}
Again using reversibility ($T\dd S = \delta Q$), this results in 
\begin{align}
    \delta Q &=  \frac{n}{2} \:\dd T = C_{JJ} \:\dd T\; (=  C_{F^{-1}} \:\dd T)\,,\\
    \delta W &= \dd H -  \frac{n}{2} \:\dd T = -\frac{1}{2}F^{-1} \dd (JJ).
\end{align}
Using the same chain of reasoning as for the previous process, this result is compared to the structure of an isobaric process in a thermodynamic canonical ensemble. In physics, such a process occurs under $\delta Q = C_p \dd T$ and $\delta W = p \dd V$, which matches the shape of the information theoretic $\delta Q =  C_{F^{-1}} \:\dd T,\;\delta W = -\frac{1}{2}F^{-1} \dd (JJ)$.

Recall that according to \cite{jaynes_information_1957}, the laws of thermodynamics are actually the laws of information itself and thus have validity outside of any physics application. 
 This Section showed that it is possible to construct \enquote{information theoretic processes} in parallel to thermodynamic processes. Hence, this result is interpreted as a confirmation of \citep{jaynes_information_1957}.

\section{Grand Gibbs ensemble}\label{sect: macrocanonical_ensemble}
The transfer of the above concepts to a grand ensemble governed by \autoref{eq:macZ}, i.e. for $\dd N \neq 0$, is straightforward: The joint partition function of $N$ non-interacting samplers in an $n$-dimensional parameter space is given by $Z[T ,J, N] = Z[T, J, N=1]^N$. Instead of controlling the number of samplers $N$, it is possible to introduce a chemical potential $\mu$, enabling the definition of the partition $\Xi[T ,J,\mu]$,
\begin{align}
    \label{eq:macZ}
    \Xi[T ,J,\mu] &= \sum_N \frac{1}{N!}Z[T ,J, N=1]^N\:\exp\left(\frac{\mu N}{T}\right) \\
    &= \exp\left(\exp\left(\frac{\mu}{T}\right)Z[T,J, N=1]\right),
\end{align}
with a Gibbs-factor $1/N!$. Effectively, the total number of samplers $N$ is replaced by $\mu$ by way of a Laplace-transform.  Importantly, \cite{herzog2023partition} 
thoroughly discuss the use of such an ensemble in MCMC sampling. The associated potential is defined as $\Omega(T,J,\mu) = - T \ln\Xi[T,J,\mu]$. Importantly, \citet{herzog2023partition} note that 
\begin{equation}
    \bra N \ket = \exp \left( \frac{\mu}{T} \right) Z[T,J, N=1] \,.
\end{equation}
In accordance with \autoref{eq: enthalpy} and \autoref{eq: entropy}, respectively, the enthalpy and entropy for the grand ensemble follow as 
\begin{align}
    H_G[T, J, \mu] &:= T^2 \partial_T \ln \Xi = \bra N \ket \left( H - \mu \right) \,, \\
    U_G [T, J, \mu] & := T^2 \partial_T \ln \Xi + \mu N = \bra N \ket H \,,\\
    S_G[T, J, \mu] &:= \partial_T (T\ln \Xi) =  \frac{\bra N \ket}{T}\left(T + H -\mu\right)\label{eq: s macro}\,.
\end{align}
with $H := H(N=1)$ the one-sampler case. 
Using \autoref{eq: s macro}, we can write
\begin{align}
    \Omega(T, J, \mu) &=  H_G[T, J, \mu] -  T S_G[T, J, \mu] \\ &= U_G  -  T S_G - \mu \bra N \ket\,,
\end{align}
which fulfills the homogeneity constraints for an extensive sampler number $\langle N \rangle$ and an intensive chemical potential analogue~$\mu$. In the Gaussian case (see \autoref{eq: Gaussian partition}), the enthalpy and entropy explicitly read  
\begin{align}
    H_G[T, J, \mu] &:= \bra N \ket \left( \frac{nT}{2} -\frac{1}{2}F^{-1}JJ - \mu \right) \,, \\
    S_G[T, J, \mu] &:= \bra N \ket \left(1 + \frac{n}{2} -\frac{1}{2T}F^{-1}JJ -\frac{\mu}{T}\right)\,.
\end{align}
In conclusion, the transfer of thermodynamic ideas to information theory naturally generalizes to a configuration with variable sampler number $\dd N \neq 0$.

\section{Relative entropies from Bayes partitions}
\label{sect_rel_entropies_from_bayes_partitions}
This section relates Bayes partitions to different concepts of relative entropy. In this, a temperature dependence of the likelihood and the prior is chosen\footnote{ The precise choice of variables does not influence the arguments in this Section. Therefore, the brackets $(\dots)$ are again used to indicate any other control variables which must not be specified in order to reach the result of Section 6.} 
\begin{equation}
    Z [T,(\dots)] = \int \dd \mu(\theta)  \:\likeli (y|\theta)^{\frac{1}{T}} \pi(\theta)^{\frac{T + 1/T}{2}}\,.\label{eq: temperature dependence relative entropies}
\end{equation}
At $T = 1$, this expression recovers the Bayes evidence, as given in \autoref{eq: evidence bayes}. The information entropy  \mbox{$S(T) =  \partial_T (T \ln Z[T,(\dots)])$} follows as
\begin{align} \label{eq:Z_KL}
     S(T)&= \ln Z[T,(\dots )] + T \partial_T \ln Z[T,(\dots)]\\ 
     &= \ln Z + \frac{1}{Z} \int \dd \mu(\theta)  \; \likeli^{\frac{1}{T}} \pi^{\frac{T + 1/T}{2}} \left(\left(T  - \frac{1}{T} \right) \frac{\ln \pi}{2} - \frac{\ln \likeli}{T} \right)  \,.
\end{align}
At this point, we set $T=1$ and employ Bayes' theorem, which leads to
\begin{align}
    S(T=1)=& \ln p(y) - \int \dd \mu(\theta) \; \frac{\likeli (y | \theta) \pi(\theta)}{p(y)} \ln \likeli (y | \theta)  \\
    =& \ln p(y) - \int \dd \mu(\theta)  \; p (\theta | y) \left(\ln p(y) + \ln \frac{p (\theta | y)}{\pi (\theta)}\right) \\
    =& - \int \dd \mu(\theta)  \; p (\theta | y) \ln \frac{p (\theta | y)}{\pi (\theta)} = - \mathrm{D}_{KL} [p (\theta | y), \pi (\theta)] \,. 
\end{align}
Thus, at $T=1$, the entropy in \autoref{eq:Z_KL} coincides with the surprise statistic mentioned above. It quantifies the amount of information obtained in an inference process, i.e. the inherent surprise of the outcome in contrast to the a priori knowledge.

Aside from the Shannon entropy, consider the class of R{\'e}nyi entropies \citep{renyi_original, Harremo_s_2006},
\begin{equation}
    S_\alpha := - \frac{1}{\alpha - 1} \ln \int \dd \mu(\theta) \; p(\theta | y)^\alpha \,,
\end{equation}
which are akin to the structure of a thermodynamic potential such as \mbox{$ G(T,(\dots)):= - T \ln \int \dd \mu(\theta)  \: (p(\theta | y) \p (y))^{1/T}$}. Here, $0<\alpha <\infty$ is a parameter $ (\alpha \neq 1)$. Using Bayes' theorem as well as an exponential likelihood and prior $\mathcal{L}\pi\sim \exp(-\Phi_J(\theta))$ (see \autoref{eq: exponential likeli}), one can write 
\begin{equation}
    p(\theta | y)_{T_0} = Z[T_0, (\dots)]^{-1}\: \exp \left(- \frac{1}{T_0} \Phi_J (\theta) \right) \,,\label{eq: T_0 posterior}
\end{equation}
for a variable temperature $T_0$. The true posterior is recovered for $T_0 = 1$. It is a well-known fact that the R{\'e}nyi entropy recovers the Shannon entropy for $\alpha = 1$. We wish to reproduce this result from the viewpoint of Bayes partitions: Substituting \autoref{eq: T_0 posterior} into the definition of the R{\'e}nyi entropy yields
\begin{align}
    \label{eq:renyi_entropy_withphi}
    S_\alpha (T_0) 
    = \frac{1}{1- \alpha } \ln \int \dd \mu(\theta) \; Z[T_0, (\dots)]^{-\alpha} \exp \left(- \frac{\alpha}{T_0} \Phi_J (\theta) \right)\,. 
\end{align}
\citet{Baez_2022} realized that the scaling with $\alpha$ and $T_0$ suggests the definition of a new temperature as $T := (T_0 / \alpha)$. With this definition, \autoref{eq:renyi_entropy_withphi} leads to
\begin{align}
    S_\alpha (T_0) &= \frac{1}{1- \alpha } \left( \ln \int \dd \mu(\theta)  \;\exp \left( - \frac{\alpha}{T_0} \Phi (\theta) \right) - \alpha \ln Z[T_0] \right)  \\
                   &= \frac{1}{1- \alpha } \left( \ln Z\left[T_0 /\alpha \right] - \alpha \ln Z[T_0] \right)  \\
                   &= \frac{1}{1 - T_0 / T} \left( \ln Z[T] - (T_0/T) \ln Z[T_0] \right)  \\
                   &= \frac{G(T)-G(T_0)}{T-T_0} \,.
\end{align}
Crucially, there exists the so-called \enquote{$q$-derivative} defined as 
\begin{equation}
    \left( \frac{\dd g}{\dd x}\right)_q := \frac{g(qx)-g(x)}{qx-x}, \; \text{for}\; q \neq 1\;\text{and}\;x \neq 0\,.
\end{equation}  
Applying this to the above result yields
\begin{equation}
    S_\alpha (T_0 = 1) = \frac{G(T_0 / \alpha) - G(T_0)}{1/ \alpha - 1} = \left( \frac{\partial G}{\partial T} \right)_{1/ \alpha} \Big|_{T_0=1}. \label{eq: hospital}
\end{equation}
Use of l'H{\^o}pital's rule shows that in the limit of $\alpha \rightarrow 1$, the $q$-derivative becomes $(\partial_x g)_{q = 1} = (\partial_x g)$. In this limit, \autoref{eq: hospital} coincides with  Shannon's information entropy~$S$.

In summary, use of the $q$-derivative allows to write the more general R{\'e}nyi entropy as a partial derivative of the Bayes partition's generating function $G = -\ln Z[T,(\dots)]$. Furthermore, this result recovers Shannon's entropy in the limit of $\alpha \to 1$, as expected (see also \citet{van_erven_renyi_2014, baez_bayesian_2014}).

\section{Effective dimensionality and model complexity}
\label{sect_effective_dimension} 
\citet{partitionfunction101} showed that for an $n$-dimensional Gaussian distribution, the number of parameters $n = \text{dim}\:\Theta$ follows from  
\begin{equation}
    \big|\partial_\beta \ln Z_G \big|_{\beta = 1} = \big|\partial^2_\beta \ln Z_G \big|_{\beta = 1} = \frac{n}{2} \, ,
    \label{eq: Gauss dimensions} 
\end{equation}
where we use the inverse temperature $\beta := 1/T$ for notational simplicity. In light of the internal energy $U=(nT/2)$ given in \autoref{eq: U for Gauss}, this result appears quite intuitive. A system governed by the entropy-maximising distribution\footnote{A Gaussian for the distributions over $\mathds{R}$ at fixed variance. We employ the Gaussian given in \autoref{eq: Gaussian likeli}.} assumes its minimum energy $U$, which is  

\begin{equation}
    U_{\text{min}}\big|_{S \;\text{max.} \leftrightarrow \; p=\text{Gauss}}(\beta=1, J =0) = \frac{n}{2}\,,
    \label{U DOFs}
\end{equation}
at $T = 1 = \beta$. In this case, $U$ is solely defined by the number of parameters $n$. This result seems to parallel an exemplary thermodynamic system such as an ideal gas, whose internal energy scales with the number of DOFs of the gas, recovering equipartition. Crucially, for a non-Gaussian distribution, the first and second derivatives in \autoref{eq: Gauss dimensions} no longer coincide. 

This discrepancy is in accordance with an information geometric observation: a Gaussian family's statistical manifold has trivial, flat geometry\footnote{In the standard KL-divergence-induced Riemannian geometry with the metric connection (the Levi-Civita connection) \citep{amari_information_2016, giesel_information_2021}.}. Due to its flatness, this manifold (and thus the structure of its statistical model) can be communicated fully by one scalar $n$. The introduction of non-Gaussianities prompts the appearance of non-trivial geometric invariants, such as scalar curvature \citep{giesel_information_2021}. Therefore, the above discrepancy suggests that the scalar produced in \autoref{eq: Gauss dimensions} is a diagnostic tool which entails more information about the statistical model (manifold) than just its dimension $n$. 

These observations motivate the definition of an effective dimension $n_\mathrm{eff}$ as a measure of deviations from the (most random, thus entropy maximising) distribution. To do this, consider $\beta \neq 1$ and once again \mbox{$Z_G[\beta,J] = \sqrt{(2\pi/\beta)^n}\:\exp\left(+\frac{\beta}{2}F^{-1}JJ\right)$}. \autoref{eq: heat gaussian} gives the heat capacity analogue for a Gaussian distribution as 
\begin{align} 
    C_{\text{Gaussian}} &:= 
    \beta^2 \partial_\beta^2 \ln Z_G \big|_{J=0} = \frac{n}{2}\label{eq: beta neq 1 } \,.
\end{align} 
\cite{schosser2024markovwalkexplorationmodel} show that the specific heat $C$ is strictly positive for Bayes partitions, which is a consequence of an inequality for logarithmic averages. This is a desirable feature for $n_\mathrm{eff}$, as it is intended to mimic a dimension. Given this result, it seems appropriate to define 
\begin{equation}
    n_\mathrm{eff} := 2\:C_{\text{non-Gaussian}}  \; > 0\;, 
    \label{eq: neff}
\end{equation} 
which is strictly positive and possibly temperature-dependent. It is expected that $n_\mathrm{eff} \leq n$, since the complexity of a sampler configuration is reduced as the appearance of higher-order cumulants marks the onset of a non-Gaussian posterior.

To inspect this, $n_\mathrm{eff}$ is derived for a weakly non-Gaussian partition function with 
\begin{equation}
    - \beta \frac{\chi^2}{2} = -\frac{\beta}{2}F_{\alpha \beta}\theta^\alpha \theta^\beta - \beta \sum_{k=3}^N \frac{1}{k!}C_{\mu_1 \dots \mu_k}\theta^{\mu_1}\dots \theta^{\mu_k}\,.
\end{equation} 
Following the treatment in \citet{partitionfunction101}, we use 
\begin{align}
    Z = \smallint \dd \mu(\theta)  \: \exp\left(- \beta\frac{\chi^2}{2} + \beta J_\alpha \theta^\alpha\right) \approxeq \: Z_G   Z_{NG} \label{eq: approximate solution for Z w. non Gauss}\,,
\end{align}
With $Z_G$ given by \autoref{eq: Gaussian partition} and 
\begin{align}
    Z_{NG} =\big[ 1-\beta \sum_{k=3} &\:\frac{1}{k!} C_{\mu_1 \dots \mu_k}\exp\left(-\frac{\beta}{2} F^{\alpha \beta} J_\alpha J_\beta\right) \nonumber \\ &\times \frac{\partial^k}{\partial J_{\mu_1} \dots \partial J_{\mu_k}} \exp\left( \frac{\beta}{2} F^{\alpha \beta}J_\alpha J_\beta \right)\bigg] \,.
\end{align}
\autoref{eq: neff} and \autoref{eq: approximate solution for Z w. non Gauss} lead to 
\begin{equation}
    n_{\mathrm{eff}} =  2 \: \beta^2 \partial_\beta^2 \ln Z_G \big|_{J=0} + 2\:  \beta^2 \partial_\beta^2 \ln Z_{NG} \big|_{J = 0}\,,
\end{equation}
which allows to write down the discrepancy $ \Delta n := n - n_{\mathrm{eff}}$, i.e. 
\begin{align}
    \Delta n &=  2 \beta^2 \sum_{\substack{k=4 \\ k\;\text{even}}} C_{\mu_1 \dots \mu_k} \frac{1}{k!} \left(2 \partial_\beta + \beta \partial_\beta^2 \right) \sum_{\pi \in \mathcal{P}} \beta ^{k/2} \prod_{\substack{s\in \pi \\|s| =2}} \prod_{i,j \in s} F^{\mu_i \mu_j}
     \\
    &=  \sum_{\substack{k=4\\ k\;\text{even}}} C_{\mu_1 \dots \mu_k} \frac{(k+2)}{2(k-1)!} \:\beta^{k/2+1} \sum_{\pi \in \mathcal{P}} \prod_{\substack{s\in \pi \\|s| =2}}  \prod_{i,j \in s} F^{\mu_i \mu_j}\,.\label{eq: neff final} 
\end{align}
Notably, $\Delta n$ is positive for $C_{\mu_1 \dots \mu_k}>0$ denoting higher-order, non-Gaussian cumulants. For $\Delta C :=  C_{\text{Gauss}} - C_{\text{non-Gauss}}$, we find 
\begin{equation}
    \Delta C =  \frac{1}{4} \sum_{\substack{k=4\\ k\;\text{even}}} \:C_{\mu_1 \dots \mu_k} \;\frac{(k+2)}{(k-1)!} \:\beta^{k/2+1} \: \sum_{\pi \in \mathcal{P}} \prod_{\substack{s\in \pi \\|s| =2}}\;  \prod_{i,j \in s} F^{\mu_i \mu_j} \,. 
\end{equation}
In this way, a difference $\Delta C$ in the analogue of the thermodynamic heat capacity can be linked to the appearance of non-Gaussian features of the posterior. From a technical viewpoint, note that only non-Gaussianities of even order contribute to $n_\mathrm{eff}$ in \autoref{eq: neff final}, as indicated in the summation. This is a relic of the approximation used to solve the integral $Z_{NG}$ before imposing $J = 0$ in $n_\mathrm{eff}$. Physically, one could argue that in a purely parabolic potential described by the Fisher-matrix $F_{\mu\nu}$, every degree of freedom would carry equal amounts of thermal energy, as required by equipartition, irrespective of affine transformations that would reshape the potential $F_{\mu\nu}\theta^\mu\theta^\nu$. As such, the dimension $n$ carries information about the number of degrees of freedom, which effectively reduces to $n_\mathrm{eff}$ smaller than $n$ in the non-Gaussian case. 

\begin{figure}
    \includegraphics[width = 0.53\textwidth]{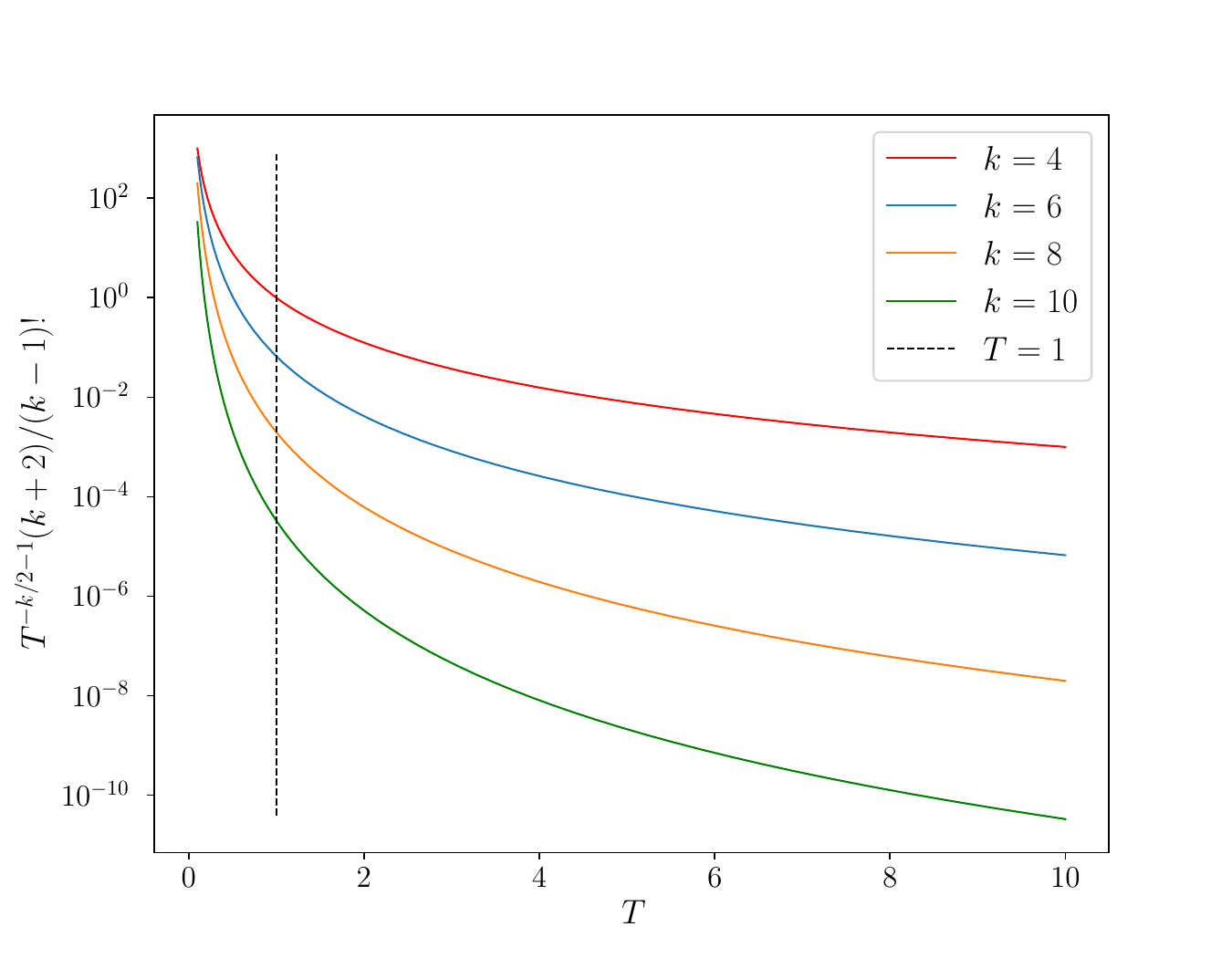}
    \caption{Decrease of $|n-n_\mathrm{eff}|$ with temperature $T$ and \emph{even} order of non-Gaussianity $k$. Here, the scaling is $T^{-k/2-1} \frac{(k+2)}{(k-1)!}$.\newline}
    \label{fig: temperature scaling}
\end{figure}

Lastly, it is worthwhile to examine the temperature scaling of the absolute difference $|\Delta n|$: for a non-Gaussianity of $k$-th order, one finds  
\begin{equation}
    |\Delta n| = |n-n_\mathrm{eff}| \sim T^{-k/2-1} \quad (k>3)\,. 
\end{equation}
This is illustrated in \autoref{fig: temperature scaling}, where the magnitude of $|n-n_{\text{eff}}|$ decreases with increasing temperature $T$. This scaling behaviour may be due to the fact that for very high $T$, the samplers are extremely mobile and swirl around in parameter space with little sensitivity to the underlying potential. Being entropy-maximising by nature, this freedom will allow the system to gradually approach the most uninformative distribution possible, i.e. a Gaussian, effectively wiping out the difference between $n_{\text{eff}}$ and $n$. Conversely, cooling or annealing to low $T$ forces the samplers to assemble close to minima of the potential, such that their distribution is indicative of the potential landscape. As a consequence, the differences between Gaussian and non-Gaussian landscapes become very pronounced at low $T$. 

Additionally, \autoref{fig: temperature scaling} shows that the higher the order of the non-Gaussianity, the more quickly the difference between $n$ and $n_\mathrm{eff}$ vanishes. Accordingly, turning up $T$ increases the gaps between different orders of non-Gaussianities. Thus, \enquote{heating} allows to effectively limit the problem to the dominant, lowest-order non-Gaussianities, whilst \enquote{cooling} evens out the weighting of the different orders, permitting higher-orders to also contribute significantly. 

\section{Cosmology Application}\label{sect_cosmology_application}
Finally, the partition function defined in \autoref{eq:canZ} and its derived formalism are applied to the distance-redshift relation of supernovae of type Ia from data compiled by \citet{riess1998observational, goobar_supernova_2011}. This data is used to constrain the matter density $\Omega_m$ and equation of state parameter $w_0$ of dark energy, and its likelihood allows to derive partition functions and information entropies (see \autoref{eq:entropy_j}).

The distance modulus $\mu$ is defined as the difference between the apparent magnitude $m$ and the absolute magnitude $M$ at a certain scale factor $a$,
\begin{equation}
    \label{eq:distance_modulus}
    \mu = m - M = 5 \log_{10}(d_L(a))+10 \,,
\end{equation}
with $d_L(a)$ the luminosity distance defined as
\begin{equation}
    d_L(a)= \frac{c}{a} \int_a^1\dd a^\prime \frac{1}{a^{\prime 2}\; H(a^\prime)}\,. \label{eq:lumi_distance}
\end{equation}
This distance measure depends on the scale factor $a$ as well as the Hubble function $H(a)$, which is given by 
\begin{equation}
    \label{eq:hubble_function}
    H(a) := H_0^{-2}\left(\Omega_m \;a^{-3}+(1-\Omega_m)\; a^{-3(1+w_0)}\right)\,.
\end{equation}
In this application, the Hubble constant is fixed to $H_0 = 70\,\text{km/s/Mpc}$ and a spatially flat spacetime is assumed. With these choices, the integral in \autoref{eq:lumi_distance} has a closed-form solution involving the hypergeometric functions \citep[see][]{arutjunjan, partitionfunction101}.
\begin{figure}
    \centering
    \includegraphics[width = 0.475\textwidth]{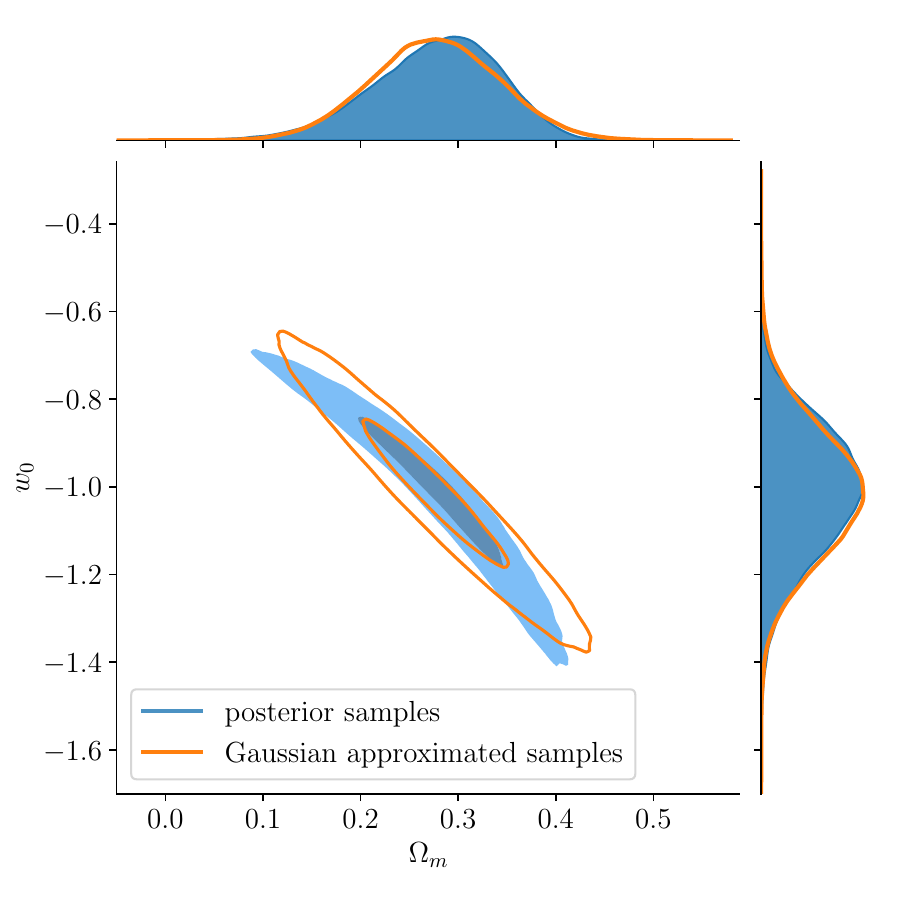}
    \caption{The Supernova posterior distribution of the matter density $\Omega_m$ and dark energy equation of state parameter $w_0$. The $2\sigma$-contour of the Gaussian approximation deviates from the true sampled posterior (blue).\newline}
    \label{fig:Supernova_posterior_Gaussian_approx}
\end{figure}
We use the Union2.1 data set \citep{suzuki2012hubble, kowalski2008improved, amanullah2010spectra} which includes SNIa distance moduli measurements at 580 redshifts. The likelihood is defined assuming Gaussian errors $\sigma_i$ and independent measurements for simplicity,
\begin{equation}
    \likeli (y \mid \Omega_m, w_0) \propto \exp\left(-\frac{1}{2}\sum_{i}\left(\frac{\mu_i - \mu(a_i | \Omega_m, w_0)}{\sigma_i}\right)^2\right)\,.
\end{equation}
A uniform prior is chosen. 
In a next step, posterior samples are obtained using the \texttt{emcee} package \citep{emcee_citation}. The resulting non-Gaussian posterior as well as a Gaussian approximation are shown in \autoref{fig:Supernova_posterior_Gaussian_approx}.

\begin{figure}
    \centering
    \includegraphics[width = 0.50\textwidth]{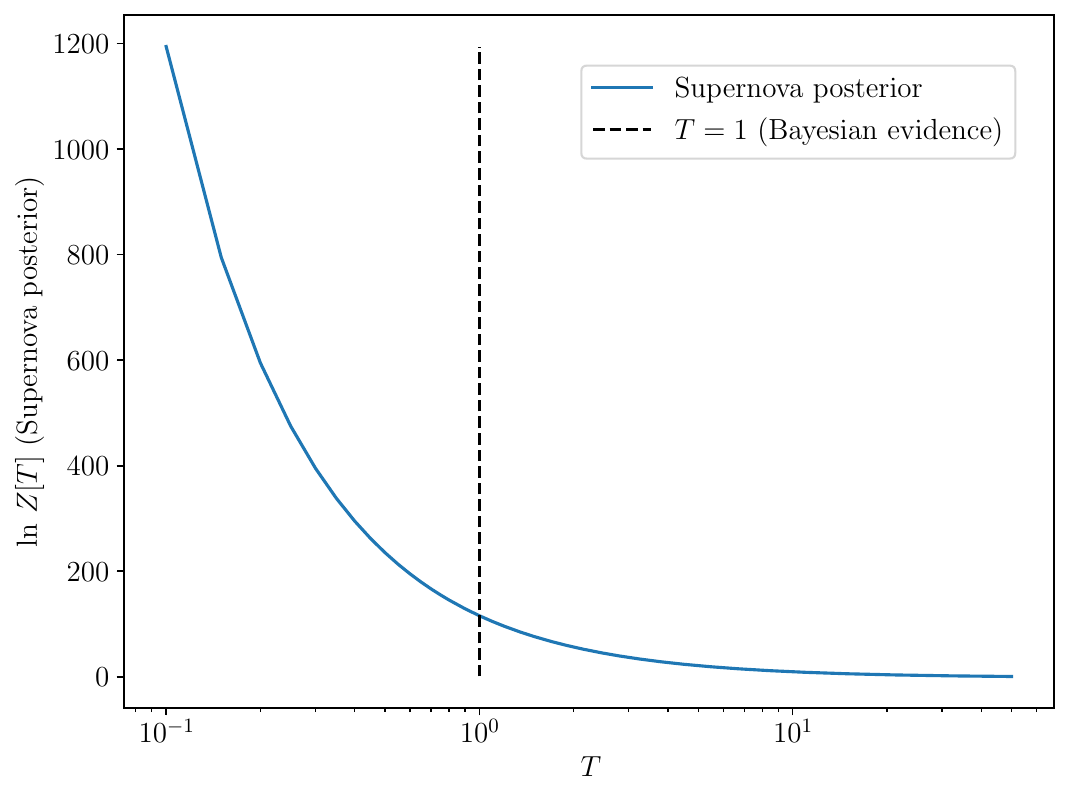}
    \caption{The generating function/ potential $\ln Z[T]$ of the Supernova posterior is given as a function of temperature. The value for $T = 1$ is marked by a vertical line, indicating the value of $\ln Z[T]$ with the Bayesian evidence $Z[T=1]$.}
    \label{fig:partition_func_temperature}
\end{figure}

\autoref{fig:partition_func_temperature} shows the partition as a function of temperature. For $T = 1$, the Bayesian evidence can be recovered from the partition function. Using Bayes' theorem, one can write the partition function as 
\begin{equation}
    \label{eq:partition_func_post}
    Z [T] = p(y)^{1/T} \smallint \dd \mu(\theta)  \: p(\theta | y)^{1/T} \,.
\end{equation}
The evidence is derived numerically using the \texttt{PyMultiNest} package \citep{pymutltinest, feroz2009multinest} which takes the likelihood $\likeli(y | \theta)$ and prior $\pi(\theta)$ as inputs. Since the Gaussian approximation shown in \autoref{fig:Supernova_posterior_Gaussian_approx} is really an approximation of the posterior samples, its likelihood is not easily accessible. Therefore, the evidence for the Gaussian approximation could not be calculated analytically by means of a density estimate of the posterior probability.

\begin{figure}
    \centering
    \includegraphics[width = 0.45\textwidth]{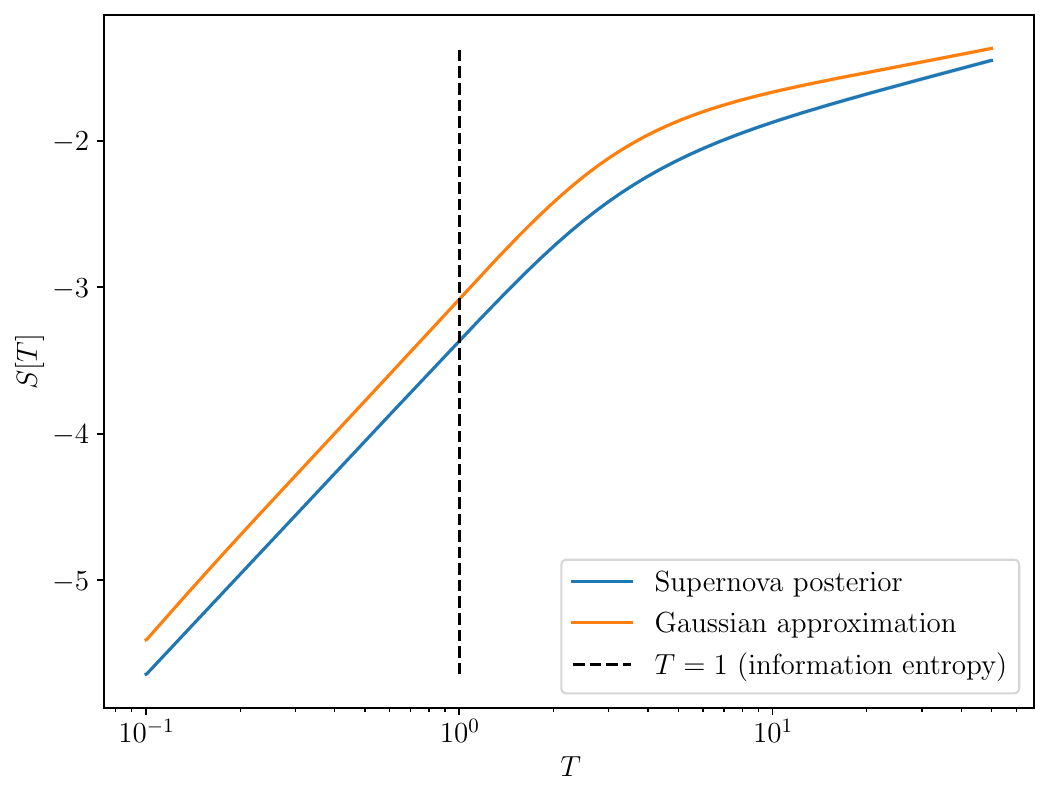}
    \caption{The thermodynamic information entropy $S[T]$ as a function of temperature $T$ is always higher for the Gaussian approximation (orange) than for its non-Gaussian counterpart (blue) - in accordance with the Gaussian distribution being entropy maximising for fixed variance. $T = 1$ marks the statistical case of information entropy.\newline}
    \label{fig: entropy_temperature}
\end{figure}

Importantly, \autoref{fig:partition_func_temperature} shows that the partition function diverges rapidly for small temperature values which stems from the exponent $1/T$ of the evidence $p(y)$ in \autoref{eq:partition_func_post}. For high temperatures, conversely, this dependence results in the convergence of $Z[T]$ to a value near zero. In this limit, the evidence becomes approximately one, whilst the integral of the posterior scaled by the inverse temperature tends to the volume of the parameter space.
Starting with \autoref{eq:partition_func_post}, we use the derived Gibbs free energy 
\begin{equation}
    \label{equ:Gibbs_free_energy_post}
    G [T] = - T \ln Z [T] = - p(y) - T \ln \smallint \dd \mu(\theta) \: p(\theta | y)^{1/T} \,.
\end{equation}
As described in \autoref{eq:entropy_j}, the temperature derivative of $G(T)$ gives the information entropy $S$. Evaluating the derivative of the Bayes partition function using finite differencing yields the entropy as a function of $T$, as shown in \autoref{fig: entropy_temperature}.
In \autoref{fig: entropy_temperature}, the entropy of the Gaussian approximation exceeds that of the posterior samples for all $T$, which illustrates the fact that the normal distribution maximises entropy at fixed variance. The entropy increases with higher $T$, which is to be expected, as the temperature effectively scales the covariance matrix. Here, higher $T$ correspond to

larger effective experimental uncertainties. \,This argument works in parallel to the $\lambda$-scaling in \autoref{eq: lambda scaling}.  The difference between the Gaussian approximation and the exact samples decreases for high $T$, meaning that the non-Gaussianities' effect reduces in this limit. Crucially, this result supports the central conclusion of Sect.~\ref{sect_effective_dimension}, which found the difference between $n$ and $n_{\text{eff}}$ to increase, as the system undergoes \enquote{heating}.
\begin{table}
    \begin{tabular}{@{}lllll@{}}
        \toprule
               & posterior samples & approximation (der.) & approximation (ana.) \\ \midrule
        $S$  & $-3.350 \pm 0.004$ & $-3.035 \pm 0.008$ & $ -3.053 \pm 0.008$\\  \bottomrule
    \end{tabular}
    \caption[Information entropy]{\raggedright Information entropy $S[T=1]$ for posterior samples and Gaussian approximation as temperature derivative of the Gibbs free energy in \autoref{equ:Gibbs_free_energy_post}. An analytical value for the Gaussian approximation is provided and falls within the two sigma interval of the value obtained by differentiation.}
    \centering
    \label{tab:entropies_Supernova}
\end{table}
For $T = 1$, the statistical information entropy can be recovered and its value is shown in \autoref{tab:entropies_Supernova}. As the covariance matrix for the Gaussian approximation is known, its entropy is calculated analytically and used as a validation of the entropy values obtained by temperature differentiation. The two entropies of the approximation do not only agree within $2\sigma$, but the results is also in accordance with that a different numerical method based on normalising flows \citep[in preparation]{partiton_flow}.

\section{summary and discussion}\label{sect_summary}
Subject of this paper was the relation between information theory applied to Bayesian inference problems on one side and the thermodynamics of sampling processes on the other. Central to this analogy is the realisation that every sampling process has an analogue in statistical mechanics. Consequently, control variables of the sampling process can be understood in parallel to thermodynamical state variables. Jeffreys covolume, which was also presented as a diagnostic marker for an inference problem's degeneracy, is used in all parameter space integrals to ensure the invariance of the volume element under reparameterizations \citep{amari_information_2016}. 
\begin{enumerate}[(i)]
\item A parametric modelling effectively scales measurement uncertainties, granting access to the intermediate stages of a Bayes update. Changes in information entropy $\dd S$ in light of new data are negative. The rate of change of the inherent \enquote{surprisal} in an outcome with a progressing Bayes update $(\dd \text{D}_{KL}/\dd \lambda)$ is a well-defined quantity. 
\item Transfer of the Jarzynski equality to information theory yields the \enquote{work} expended in a Bayes update as a function of the log-likelihood, i.e. $W = -\lambda \ln \mathcal{L}$, at the current state $\lambda$ of the update step. 
\item The transfer of statistical mechanics vocabulary to statistical inference can occur through Bayes partitions. An information theoretic analgoue of the Guggenheim scheme for a Gaussian likelihood and uninformative prior suggests the inverse Fisher information $F^{\mu\nu}$ as an extensive measure of \enquote{system size}. The structure of the internal energy $U$ and heat capacity analogue $C$ match the respective results for a monoatomic ideal gas in statistical physics. Interestingly, the number of parameter space dimensions $n$ assumes the position held by the number of internal DOFs in thermodynamics. 
\item An inspection of Bayes updates as thermodynamic processes shows that changes in \enquote{work} $\delta W$ and \enquote{heat} $\delta Q$ can be formulated meaningfully for a sampler ensemble in parameter space. The close resemblence to physics processes is interpreted as a high-level demonstration of the fundamental result by \citet{jaynes_information_1957}. 
\item A specific temperature dependence of the Bayes partition (\autoref{eq: temperature dependence relative entropies}) leads to the \emph{surprise statistics}, i.e. the relative entropy between prior and posterior. Bayes partitions successfully recover Shannon's information entropy from Rényi entropies by way of $q$-derivatives. 
\item An effective dimension $n_{\text{eff}}$ (or equivalently an effective heat capacity $C_{\text{eff}}$) serves as an indicator of deviations from Gaussianity. Crucially, $n_{\text{eff}}\:(\leq n)$ is a measure of system complexity, which reduces as higher-order correlations appear. The effective dimension is maximized for the entropy-maximizing distribution, a Gaussian, in which case $n_{\text{eff}} = n$ and $C = (n/2)$. In the event of heating (increase in $T$), the difference $\Delta n = |n_{\text{eff}}-n|$ gradually vanishes, suggesting that overly mobile samplers lack sensitivity to detect differences in the underlying potential landscapes. Conversely, $\Delta n$ increases with cooling, as the potentials imprint on the sampler distribution in much more stringency. The higher the order of non-Gaussianity, the more quickly its influence on the sampler ensemble drops off with $T$.
\item The formalism is illustrated on the inference of the matter density parameter $\Omega_m$ and the dark energy equation of state parameter $w_0$ using a data set of SNIa distance moduli. The Gaussian approximation of the posterior deviates from the exact posterior (sampled using \texttt{emcee}),a s anticipated. At $T = 1$, the Bayes partition coincides with the evidence that was determined using \texttt{PyMultiNest}. In the limit of high $T$, the Bayes partition approaches a constant value. Finite differencing with respect to $T$ allows to access the information entropy $S$ of the Gaussian, as well as of the exact posterior. The results agree with the entropy-maximization provided by a Gaussian distribution. Crucially, $\Delta S$ between the Gaussian and non-Gaussian distribution decreases with heating. Crucially, this finding separately confirms the result of Sect.~\ref{sect_effective_dimension}, which found $\Delta n$ to decrease with increasing $T$.
\end{enumerate}

\section*{Acknowledgements}

\paragraph{Funding information}
This work was supported by the Deutsche Forschungsgemeinschaft (DFG, German
Research Foundation) under Germany's Excellence Strategy EXC 2181/1 - 390900948
(the Heidelberg STRUCTURES Excellence Cluster). RMK acknowledges funding of the
Stiftung der Deutschen Wirtschaft (Foundation of German Business) with funds
from the Begabtenförderung of the BMBF (Federal Ministry of Education and
Research's scholarship programme for gifted students). HvC
 is supported by the Konrad Zuse School of Excellence in Learning
and Intelligent Systems (ELIZA) through the DAAD programme Konrad Zuse Schools
of Excellence in Artificial Intelligence, sponsored by the Federal Ministry of
Education and Research. 
We acknowledge the usage of the AI-clusters {\em Tom} and {\em Jerry} funded by
the Field of Focus 2 of Heidelberg University.

\paragraph{Data availability}
No new data resulted from this study.

\bibliography{references}

\end{document}